\documentclass[trackchanges, onecolumn]{aastex631}

\usepackage{wrapfig}
\usepackage{rotating}
\usepackage{hyperref}

\shorttitle{U Gem Nova Shells}
\shortauthors{Shara, Lanzetta, Doshi et al.}

\begin{document}


\title[U Gem Nova Shells]{Two Predicted, Concentric Nova Shells Surround the Prototype Dwarf Nova U Geminorum}

\correspondingauthor{Michael M. Shara  mshara@amnh.org}

\author[0000-0003-0155-2539]{Michael M. Shara}
\affiliation{Department of Astrophysics, American Museum of Natural History, New York, NY 10024, USA}

\author[0000-0001-6906-7594]{Kenneth M. Lanzetta}
\affiliation{Department of Physics and Astronomy, Stony Brook University, Stony Brook, NY 11794-3800, USA}

\author[0000-0000-0000-0000]{Aayushi Doshi}
\affiliation{Department of Astrophysics, American Museum of Natural History, New York, NY 10024, USA}

\author[0000-0002-3361-2893]{Alexandra Masegian}
\affiliation{Department of Astronomy, Columbia University, New York, NY 10027, USA}

\author[0009-0006-3617-1356]{Stefan Gromoll}
\affiliation{Amazon Web Services, 410 Terry Ave. N, Seattle, WA 98109, USA}

\author[0000-0002-0023-0485]{Yael Hillman}
\affil{Dept. of Physics, Azrieli College of Engineering, Jerusalem 9103501, Israel}

\author[0000-0001-5384-7545]{Susanne M. Hoffmann}
\affil{University of Science and Technology of China, Dept. of the History of Science and Scientific Archaeology, Hefei 230026, China}

\affil{Friedrich Schiller University of Jena, Faculty of Mathematics and Computer Science, FUSION Group, 07737 Jena, Germany}

\author[0000-0002-9821-2911]{David Valls-Gabaud}
\affil{Observatoire de Paris, LUX, CNRS UMR 8112, 61 Avenue de l’Observatoire, 75014 Paris, France}

\author[0000-0001-7796-1756]{Frederick M. Walter}
\affil{Department of Physics and Astronomy, Stony Brook University, Stony Brook, NY 11794-3800, USA}

\author[0000-0002-0004-9360]{John K. Webb}
\affil{Institute of Astronomy, University of Cambridge, Madingley Road, Cambridge CB3 0HA, United Kingdom}

\begin{abstract}
In the 170 years since its discovery, U Gem has been intensively studied as {\it the} prototypical cataclysmic binary star. Its massive white dwarf (WD) ($\sim$ 1.2 $M_{\odot}$) is cannibalizing its $\sim$ 0.42 $M_{\odot}$ red dwarf companion. The WD's resulting accreted hydrogen-rich envelope has previously undergone thermonuclear runaways seen as nova eruptions. These $\sim$ weekslong transient events brighten U Gem to $m_{V} \sim$ -3 and eject the accreted envelope ($\sim$ $10^{-5} M_{\odot}$) at thousands of km/s. Ultraviolet observations show that the accreted envelope of U Gem's WD is greatly enhanced in nitrogen and depleted in carbon relative to solar composition, and that the WD's effective temperature is much hotter than that of most cataclysmic binary WDs. These suggest that U Gem underwent a classical nova eruption quite recently and should therefore still be surrounded by the ejecta of that event. Modeling of U Gem predicts a nova event $\sim$ 1000 years ago, and the existence of {\it two} concentric shells, each of order 1 degree in size surrounding U Gem. We obtained deep narrowband H$\alpha$ imaging of U Gem with the Condor Array Telescope. The two concentric H$\alpha$-bright shells that we find, centered on U Gem, may be the first-ever predicted old nova ejecta. A transient ``guest star" in the asterism Shuiwei, recorded by Chinese imperial astrologers in November 829 CE is consistent with the timing and location of U Gem's last predicted nova eruption, the transient's absence in Japanese and Korean records weighs against this suggestion.
\end{abstract}


\section{Introduction} \label{sec:intro}

\subsection{Dwarf Novae (DNe)}
Dwarf novae (DNe) are cataclysmic binary systems, comprising a white dwarf (WD) and a low-mass companion star that fills its Roche lobe \citep{Kraft1962,Warner1995}. Material from the companion flows through the inner Lagrangian point of the binary system, forming an accretion disk that orbits the white dwarf. A thermal instability of the accretion disk (typically every few weeks to years) leads to the rapid accretion of much of the disk matter onto the WD \citep{Osaki1974,Osaki1996}. The resulting release of gravitational potential energy results in the $\sim$ 2-6 magnitude brightenings of days to a few weeks duration that characterize dwarf nova outbursts \citep{Smak1971}. 

\subsection{U Geminorum}

U Geminorum (U Gem) was the first dwarf nova ever discovered. 
John Russell Hind \citep{Hind1856} described his discovery as follows: 
``On the evening of December 15th, 1855, I remarked...an object shining as a
star of the ninth magnitude, with a very blue planetary light,
which I have never seen before during the five years that my
attention has been directed to this quarter of the heavens. On
the next fine night, Dec. 18, it was certainly fainter than on the
15th by half a magnitude or more. Since that date I have not
had an opportunity of examining it till last evening, January 10th,
when its brightness was not greater than that of stars of the
twelfth magnitude. It is evidently a variable star of a very interesting description, 
inasmuch as the minimum brightness appears to extend over a great part of the whole period, 
contrary to what happens with Algol and S Cancri."

As one of the nearest ($\sim$ 93 parsecs) \citep{Gaia2023} and brightest-at-maximum-light DNe, U Gem has been intensively observed and studied since its discovery 170 years ago \citep{Cook1987}. U Gem exhibits outbursts spaced, on average, 118 days apart. It brightens from $m_{V}$ $\sim 14.9$ to $\sim 8.2$ mag for an average outburst time of $\sim$ 12 days \citep{Szkody1984,Ritter2003}. It has been detected in the radio \citep{Coppejans2016}, infrared \citep{Harrison2000}, ultraviolet \citep{Fabbiano1981}, extreme ultraviolet \citep{Long1995}, and X-ray \citep{Mason1978} wavelengths. Multiwavelength observations have yielded a well-documented set of U Gem binary parameters and properties, some of which we now list and will use in Section 2. 

U Gem is a partially eclipsing binary with an orbital period of 0.1769061911 days (4 hr 14.745 minutes), a mass ratio q = 0.35 $\pm$ 0.05, a WD mass $M_{1}$ $\sim$ 1.2 $M_{\odot}$ and a red dwarf mass of 0.41 - 0.42 $M_{\odot}$ \citep{Echevarria2007}. In quiescence, shortly after a DN outburst, the WD displays a photospheric temperature of 41,500 Kelvin \citep{Godon2017}, cooling to $\sim$ 27,000 Kelvin within several months \citep{Long1994,Long1995,Sion1998,Godon2017}. This is still much hotter than the temperatures displayed by the white dwarfs of most cataclysmic binaries (typically 10,000 -15,000 Kelvin \citep{Pala2022}), suggesting that the WD of U Gem may have undergone a nova thermonuclear runaway relatively recently (see below).

A second piece of evidence suggesting a recent nova in U Gem is the very non-solar metal abundances in its WD's photosphere \citep{Long1999,Godon2017}. These show evidence of significant CNO-cycle processing, particularly the WD photosphere's subsolar carbon and $>$10X solar nitrogen abundances. These abundances strongly suggest that the matter currently being transferred from the red dwarf onto the WD in U Gem has undergone hot CNO-cycle burning \citep{Gansicke2003,Godon2011,Godon2023}. A plausible explanation is that the deduced enrichment of the outer layers of the convective red dwarf envelope occurred over several Gyr, during $\sim$ $10^{5}$ nova eruptions. Each such eruption transferred hot CNO-cycle-processed matter from the WD to its companion \citep{Figueira2018,Figueira2025}, which is now being returned to the WD. 


\subsection{Novae and U Gem}
A nova eruption is caused by a thermonuclear runaway in the envelope of hydrogen-rich matter \citep{starrfield1972,prialnik1979,Yaron2005} that has been accreted from a binary companion onto a WD. Novae achieve peak luminosities of $10^{4}$ to $10^{6}L_{\odot}$ for days to years, before extinguishing themselves by ejecting their accreted envelopes \citep{prialnik1978}. All novae recur thousands of times \citep{Ford1978}, with inter-eruption periods typically ranging from millennia to Myr over their multi-Gyr lifetimes \citep{Shara2018,Hillman2020}. U Gem is no exception. Since hydrogen-rich matter is being accreted (episodically) onto the WD of U Gem at a rate of $\sim$ $10^{-10}$ $M_{\odot}$/yr \citep{Sion1998,Froning2001}, its inter-eruption timescale (given its 1.1 - 1.2 $M_{\odot}$ WD) is $\sim$ 10-100 kyr \citep{Yaron2005}. Given both its very hot WD and highly CNO-processed material on its RD, U Gem is a good candidate for being the first (non-Z Cam type) dwarf nova \citep{Shara2012a,Shara2012b} whose last nova eruption occurred quite recently. If so, its ejecta may not have had time to dissipate and fade to the point of being undetectable.

 Models of U Gem nova eruptions and their resultant predicted shells are discussed in Section 2. In Section 3 we describe the observational data of this paper, motivated by the Section 2 prediction of U Gem shells. Section 4 presents images of the newly discovered nebulosities surrounding U Gem. In Section 5, we note that Chinese Imperial astrologers may have recorded a nova eruption of U Gem in 829 CE. We briefly summarize our results in Section 6.

\section{The Prediction of Concentric Nova Shells Surrounding U Gem}

\subsection{The Code}

We can only crudely estimate the inter-eruption timescale (10--100 kyr) for U Gem from the grid of nova models in \citep{Yaron2005}. We therefore used our self-consistent nova binary evolution code \citep{Hillman2020} to better estimate the time since U Gem's last eruption and the characteristics of that eruption. The code simulates Roche Lobe Overflow (RLOF) evolution in CVs, which leads to periodic nova eruptions, while self-consistently calculating the mass transfer rate at each time step. In addition to accounting for angular momentum (J) losses due to magnetic braking and gravitational radiation, the code tracks the redistribution of J caused by mass transfer from the red dwarf (RD) to the white dwarf (WD) at each timestep, and the loss of J due to mass ejection from the system during a nova eruption. These changes drive the orbital period and binary separation, which in turn control the mass transfer rate dM/dt. The code tracks how dM/dt increases with decreasing separation and decreases with increasing separation during each timestep of each nova cycle. During a nova eruption, the RD is strongly irradiated, so its envelope temporarily leaves thermal equilibrium (typically for a few centuries \citep{Kovetz1988}). The resulting expansion increases the extent of RLOF, leading to a higher accretion rate after an eruption \citep{Hillman2020,Hillman2021}. 

\subsection{Initial Conditions and Runtime}

We began our simulation with initial binary masses and an orbital period consistent with those observed for U Gem, namely, a CO WD mass of $1.2 M_\odot$ with a core temperature of 10 MK, and a companion RD of mass $0.42M_\odot$. The system began at an orbital period of $\sim4.24$ hours and was allowed to evolve as described in the previous section. After modest fluctuations in the first few eruptions' inter-eruption times, peak luminosities, and ejection velocities, subsequent eruptions were all very similar (see Figure 1). The simulation continued uninterrupted for $\sim8\times10^5$ years, during which changes in the system parameters (i.e., the WD and RD masses and the orbital period) were negligible. 

\subsection{Simulation Predictions}
\subsubsection{Two Shells}

Figure 1 shows that the simulated WD underwent $\sim35$ nova eruptions separated by an average of $2.3\times10^4$ years. During each eruption it brightened $\sim$ 100,000-fold to a peak luminosity of $\sim$ $10^{5} L_{\odot}$, before declining back to its observed quiescence luminosity $\sim$ $L_{\odot}$. Figure 2 plots the calculated ejection velocities during one of those nova outbursts (all of which are very similar). In the first few hours of the nova eruption, the thermonuclear runaway drives a shock into the WD's outer envelope, ejecting matter at terminal speeds up to 2,500 km/s. The resulting expansion and cooling of the envelope decreases its rate of energy production so that, for the next $\sim$ five days, only enough energy is produced to eject matter at 100-200 km/s. The largely depleted (via ejection) hydrogen-rich envelope contracts enough for a final strong flash to occur on day 6, accompanied by another pulse of rapidly ejected ($\sim$ 2000 km/s) material. This last, high-speed ejected matter should catch up to and merge with the slow ejecta just a few hours after it is produced. After this final ejection phase, the envelope is exhausted, which terminates the nova eruption \citep{prialnik1978}. The net (predicted) result is a {\it pair} of concentric nova shells surrounding U Gem. The ratio of their sizes depends on the mass and velocity distribution of the outer (high-speed) ejecta, the density and distribution of the ISM surrounding U Gem just before the nova eruption (e.g., was it depleted by a previous nova eruption?), and the speed of the slow ejecta.

Two effects not noted above will also affect the sizes and morphologies of U Gem ejecta. First, while the simulation described above models time-variable accretion, the resulting feedback history, and the thermonuclear runaway in great detail, it is 1D. Novae eject their envelopes asymmetrically, in faster-moving polar ``fans'' and slower-moving equatorial rings \citep{Gill2000}. In addition, matter ejected at speeds comparable to or slower than the orbital velocity of the donor star (typically 100-300 km/s) will accelerate as the donor moves supersonically through the WD-ejected wind \citep{Lloyd1997}. These two effects suggest that the morphology of the ejecta of U Gem nova may be more complex than two neatly nested, spherical shells. Nonetheless, the predictions of both fast and slow ejecta suggest searching not only for a single nova shell surrounding U Gem, but also for two roughly concentric structures.

\subsubsection{Time Since the Last Nova Eruption}

Ultraviolet observations of U Gem show that its WD's effective temperature, Teff, cools monotonically in the months after a dwarf nova accretion event. The longest ($\sim$ 4 months) gaps between DN outbursts show Teff decreasing to 27,000 Kelvin. Figure 3 is a plot of the predicted WD effective temperature as a function of time after a U Gem nova eruption. The cooling WD requires $\sim$ 800 yrs to reach an effective temperature of $\sim$ 27,000 K. Since repeated dwarf nova accretion events inject heat into the outer layers of U Gem's WD, 800 yrs is a lower limit on the time since the last U Gem nova eruption. Nonetheless, the high effective temperature of U Gem's WD relative to those of almost all other WD CVs \citep{Pala2022}, which are also undergoing dwarf nova eruptions, suggests a nova event much more recently than the $\sim$ 23,000 yr inter-nova event rate for U Gem predicted by our simulations.

\section{The Data}\label{sec:datasets}

To test the prediction that U Gem underwent a relatively recent ($<<$ 23,000 yr) nova eruption, and that it may be surrounded by two shells of ejecta, we imaged it with the Condor Array Telescope \citep{Lanzetta2023}. Condor comprises six fast (f/5) apochromatic refractors with 180 mm objectives. Each refractor is equipped with a 9576 x 6388-pixel CMOS camera covering 2.3 x 1.5 $\mathrm{deg^{2}}$, with a plate scale of 0.86 arcsec/pixel. \citet {Lanzetta2023} describe the telescope, image-signature removal (flat-fielding, etc.), and astrometry.

We imaged through three narrowband filters, each with a 3 nm FWHM, with the following central wavelengths (CWL): [OIII] 500.7 nm, H$\alpha$ 656.3 nm, and [SII] 671.6 nm. In addition, it was imaged through a broadband Luminance filter with a FWHM of 280 nm and a CWL of 550 nm. 

The telescope was dithered by a random offset of $\sim$ 15 arcmin between each 
exposure. Because different filters were used on different telescopes at different 
times, we define the “reach” of an observation obtained by Condor observations as 
the product of the total objective area and the total exposure time devoted to the 
observation. As Condor consists of six individual telescopes, each of objective 
area 0.0254 $m^{2}$, a one-second exposure with one telescope of the array yields a reach 
of 0.0254 $m^{2}$ s, and a one-second exposure with the entire array (i.e. with all six 
telescopes) yields a reach of 6 x 0.0254 $m^{2}$ s = 0.153 $m^{2}$ s. Table 1 presents the observation dates and image reaches for U Gem. We often used three or four telescopes simultaneously with H$\alpha$ filters, which provided deeper reach in that filter.

\begin{table}[ht]
\centering
\begin{tabular}{lcccc}
\multicolumn{4}{c}{{\bf Table 1:}  Details of Observations} \\

\multicolumn{1}{c}{Condor Field U GEM} &\multicolumn{1}{c}{J2000} & \multicolumn{1}{c}{R.A. = 07:55:05} &\multicolumn{1}{c}{Dec = +22:00:05}\\
\hline
\hline

\multicolumn{1}{l}{Filter} & \multicolumn{1}{c}{Start Date} & \multicolumn{1}{c}{End Date} &\multicolumn{1}{c}{Reach ($m^{2}$s)} \\
\hline
\multicolumn{1}{l}{Luminance 550.0 nm}  & 2023-10-05 & 2023-12-31& 975 \\
\multicolumn{1}{l}{[OIII] 500.7 nm}    & 2023-10-06 & 2024-02-04 & 3673 \\
\multicolumn{1}{l}{H$\alpha$ 656.3 nm}   & 2023-10-06 & 2023-12-31 & 12192 \\
\multicolumn{1}{l} {[SII] 671.6 nm}     & 2024-02-04 & 2024-02-04 & 61 \\

\hline
\end{tabular}
\end{table}

\section{U Gem Images}\label{}
\subsection{Narrowband and Luminance Images}

Figure 4 shows the three U Gem images taken through the narrowband filters noted above, as well as the broadband Luminance filter image. Figure 5 again shows the Luminance image, now with the difference images formed from it and the narrowband images. Nebulous structures are clearly seen surrounding U Gem in H$\alpha$. A ring of [OIII] emission may be present, but it is near the detection limit, while the [SII] image (see Table 1) is too shallow to draw any conclusions.

 The brightest areas in the H$\alpha$ image are radiating with a surface brightness $\sim$ $10^{-16}\,\mathrm{erg/s/cm^{2}/arcsec^{2}}$. The H$\alpha$ surface brightness 1-sigma uncertainty over one 0.86" pixel is 3.1 x $10^{-17}\,\mathrm{erg/s/cm^{2}/arcsec^{2}}$, corresponding to a surface brightness 3-sigma uncertainty over a 10 x 10 $\mathrm{arcsec^{2}}$ region of 8.0 x $10^{-18}$\, $\mathrm{erg/s/cm^{2}/arcsec^{2}}$.

\subsection{Net H$\alpha$ brightness contours}

To create star-subtracted images, we processed the Luminance and H$\alpha$ Condor coadded mosaics separately before subtraction. We loaded the input images from the Condor coadded FITS files, binned them, and median-background-subtracted them to establish a consistent baseline for masking across both filters. We then generated a conservative automatic mask to identify compact stellar residuals, which we modestly expanded to include the immediate wings of detected sources. We additionally masked bright stars, saturated cores, and extended residual structures not fully captured by the automatic mask in the binned-pixel frame. We treated these manually masked regions as contaminated areas rather than locations where we could reliably recover the underlying astrophysical signal.

After defining the masks, we cleaned the Luminance and H$\alpha$ mosaics separately using local median replacement, which provides a stable treatment for the crowded stellar field while minimizing large-scale interpolation artifacts. We then generated the cleaned H$\alpha$-L image only after independently processing both input mosaics. For the most severe bright-star residuals, a local annulus-based repair was applied to the final difference image for display, using background statistics to suppress residual stellar artifacts without imposing or forcing shell-like structure through the contaminated regions.

We created contours from a contour-optimized version of the cleaned H$\alpha$-L image. To extract contours, we excluded the outer 100–200 pixels to prevent edge-related artifacts and omitted manually masked bright-star regions from the contour map. We then derived a single contour level from a smoothed version of this contour image to highlight the main large-scale morphology of the enhanced H$\alpha$-L emission. Figure 6 shows the result. 

The contour in Figure 6 serves as a morphological guide to the central annular and surrounding horseshoe-shaped emission structures and excludes regions dominated by bright stars, image edges, or known instrumental artifacts. It does not prove that U Gem is surrounded by one or more nova shells, but it does demonstrate i) a central H$\alpha$-bright region $\sim$ 40' x 20' (1.1 x 0.55 pc at the 93 pc distance of U Gem) in size, ii) a semicircular arc $\sim$ 1.5 degrees (2.4 pc) in size to the East of the central structure, and (iii) a bi-lobed, inverted U-shaped structure $\sim$ 1 degree (1.6 pc) in size to the West of the central structure. 

The largest nova shells of 20th-century novae are $\sim$ 0.2 pc in size or smaller \citep{Gill2000,Shara2012c}, while that of $\sim$ 330-year-old AT Cnc is $\sim$ 0.35 pc in diameter \citep{Shara2012a}; few older novae are known \citep{Shara2012b}. The most extensive compilation of nova shells is that of \citet{Santamaria2025}, who showed that novae, on average, grow in radius at a rate $\sim$ 0.0725 pc/century. If the H$\alpha$ structures of Figure 6 are, indeed, nova shells ejected from U Gem, their sizes are consistent with eruptions in the past few thousand years. Detailed spectroscopy will be needed to determine the kinematics, density and composition of the nebulosities of Figure 6, so that these can be compared to those of known nova ejecta. 

\section{An Historical Counterpart?}

Figure 7 shows a map of the Chinese asterisms \citep{Yang2025} surrounding U Gem. U Gem is closer to Shuiwei (the Water Level) than to any other asterism.

A quote from the "History of the Tang Dynasty"\footnote{Compiled 1043-1060 by Ouyang Xiu and Song Qi} is \citep{Yang2025}:
"Emperor Wenzong of Tang, 3rd year of the Taihe reign period, 10th month. A guest star appeared in Shuiwei." (Xin Tang shu, Tianwen zhi, ch. 32). This corresponds to November 829 CE. While no details were recorded about the brightness or interval of visibility of this guest star, it is the only mention of a transient in Shuiwei during the time period 500 BCE through 1600 CE \citep{Xu2000}.

We regard this particular record as relatively reliable, as it appears only in the "New Book of Tang" written in the 11th century, and not in the earlier "Old Book of Tang". In the old tradition, scholars recorded phenomena only when they believed they were relevant to omens. In contrast, the author of {\it Xin Tang Shu} (the New Book of Tang), Ouyang Xiu, was skeptical about the predictive significance of celestial events and presented a list of celestial phenomena without tying them to specific portents or earthly events. It thus includes more (and perhaps more unbiased/ unframed) observations than its predecessor. 

Unfortunately, no known confirmatory records from Japan or Korea \citep{Park2022} exist, even though observers were recording celestial phenomena in both locations in 829 CE \citep{Ho1962}. U Gem would have been easily visible from both countries after $\sim$ midnight in November, reaching visual magnitude $\sim$ -2 to -3. The lack of any confirmation of what must have been a remarkably bright transient suggests that, if real, it might have been a fireball. Hopefully independent confirmation and/or other details will emerge from future research of ancient Asian records. 

\section{Conclusions}

The very hot WD of U Gem, and the distinctly non-solar abundances of its WD's accreted envelope, suggest that this prototypical dwarf nova has undergone a relatively recent classical nova eruption. Self-consistent simulations of multiple nova eruptions in the U Gem binary predict a fast and a slow wind in each eruption, hence the existence of two concentric shells of ejecta surrounding U Gem. Deep narrowband Condor H$\alpha$ imaging of the environs of U Gem reveals two concentric shells, as predicted. A transient guest star recorded by Tang Dynasty astrologers in November 829 CE, at roughly the location of U Gem, is consistent with the timing of its last nova eruption, but no details of the brightness or duration of the event are known to exist in Chinese, Japanese, or Korean records of celestial transients. 

\section{Acknowledgments}
KML, MMS, and AD acknowledge the support of National Science Foundation Grants 1910001, 2107954, 2108234, 2407763, and 2407764. We gratefully acknowledge Sheri Henriksen's generous support for the Condor Array Telescope. This work uses data from the European Space Agency (ESA) mission {\it Gaia} (\url{https://www.cosmos.esa.int/gaia}), processed by the {\it Gaia} Data Processing and Analysis Consortium (DPAC, \url{https://www.cosmos.esa.int/web/gaia/dpac/consortium}). Funding for the DPAC has been provided by national institutions, in particular those participating in the {\it Gaia} Multilateral Agreement. We thank Bo-shun Yang for valuable conversations and suggestions regarding the Shuiwei transient of 829 CE.

\bibliography{UGem}
\bibliographystyle{aasjournal}

\newpage
\begin{figure*}[h!]
    \centering
    \includegraphics[width=7in]{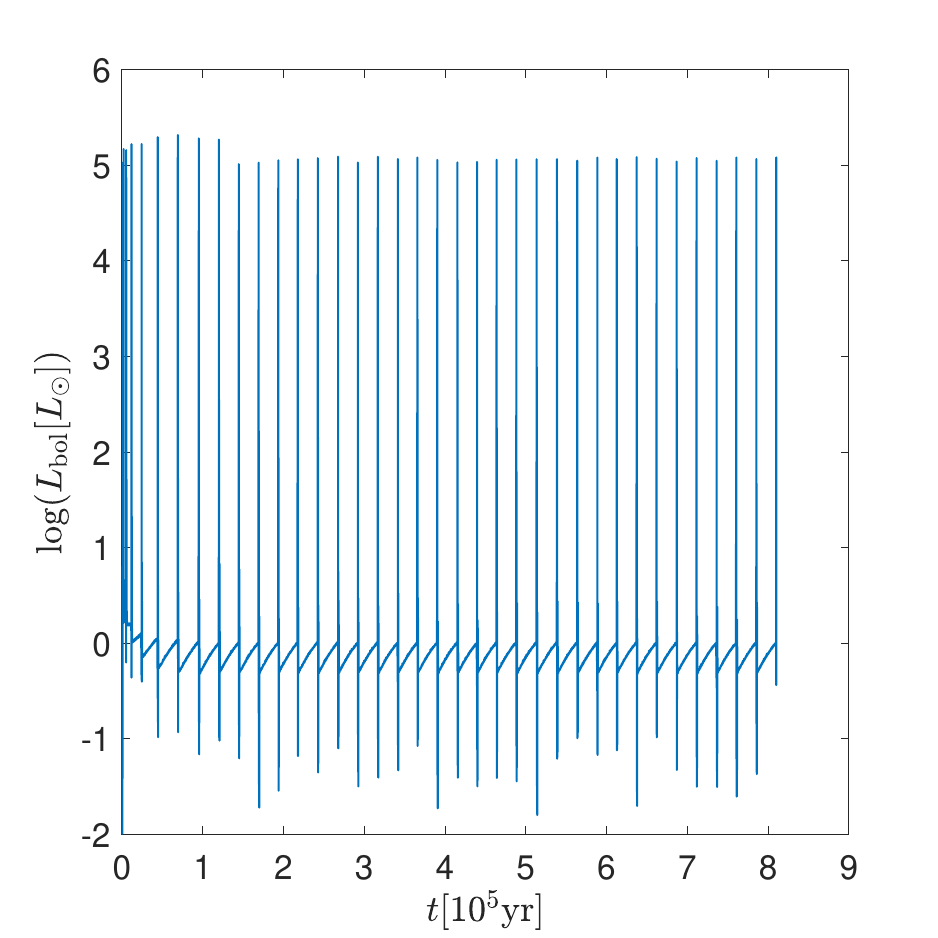}
    \caption{The luminosity as a function of time predicted for the cataclysmic binary U Geminorum by the self-consistent numerical simulation described in Section 2. After a few irregularly spaced (in time) eruptions, the binary reaches a quasi-steady state with nova thermonuclear eruptions producing peak luminosities $\sim$ $10^{5} L_{\odot}$ every $\sim$ 23,000 yrs.}
\end{figure*}

\newpage
\begin{figure*}[h!]
    \centering
    \includegraphics[width=7in]{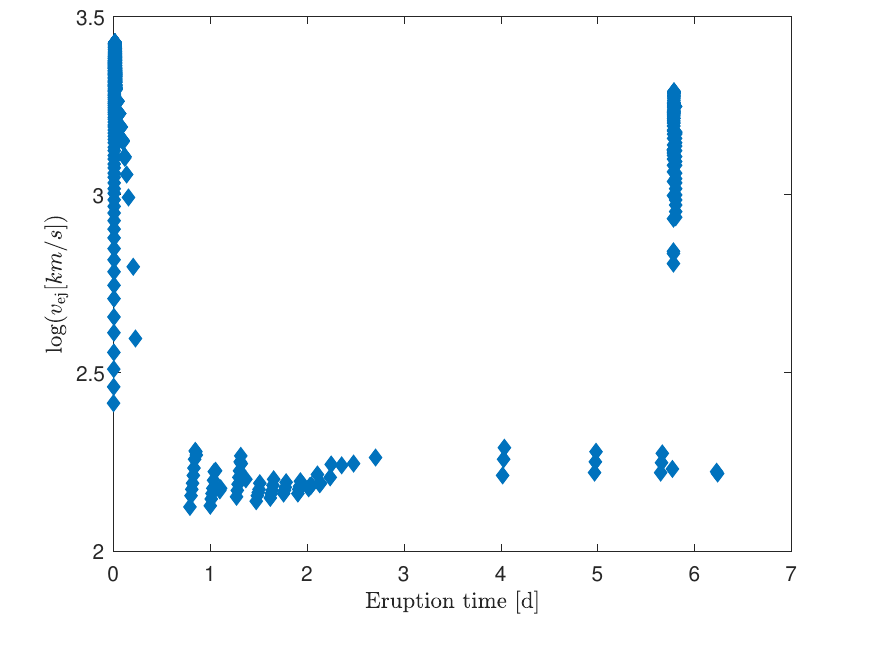}
    \caption{The velocities of thin shells ejected from U Gem during a nova outburst, as predicted by the simulation described in Section 2. The initial thermonuclear runaway leads to a fast wind, with maximum speeds of $\sim$ 2500 km/s, lasting a few hours. This is followed by $\sim$ six days of mass ejection at much slower speeds ($\sim$ 200 km/s); and a final short burst of high-speed wind that will catch up to and merge with the slow wind. The model predicts two roughly concentric shells of ejecta surrounding U Gem, produced by its most recent nova outburst.}
\end{figure*}

\newpage

\newpage
\begin{figure*}[h!]
    \centering
    \includegraphics[width=7in]{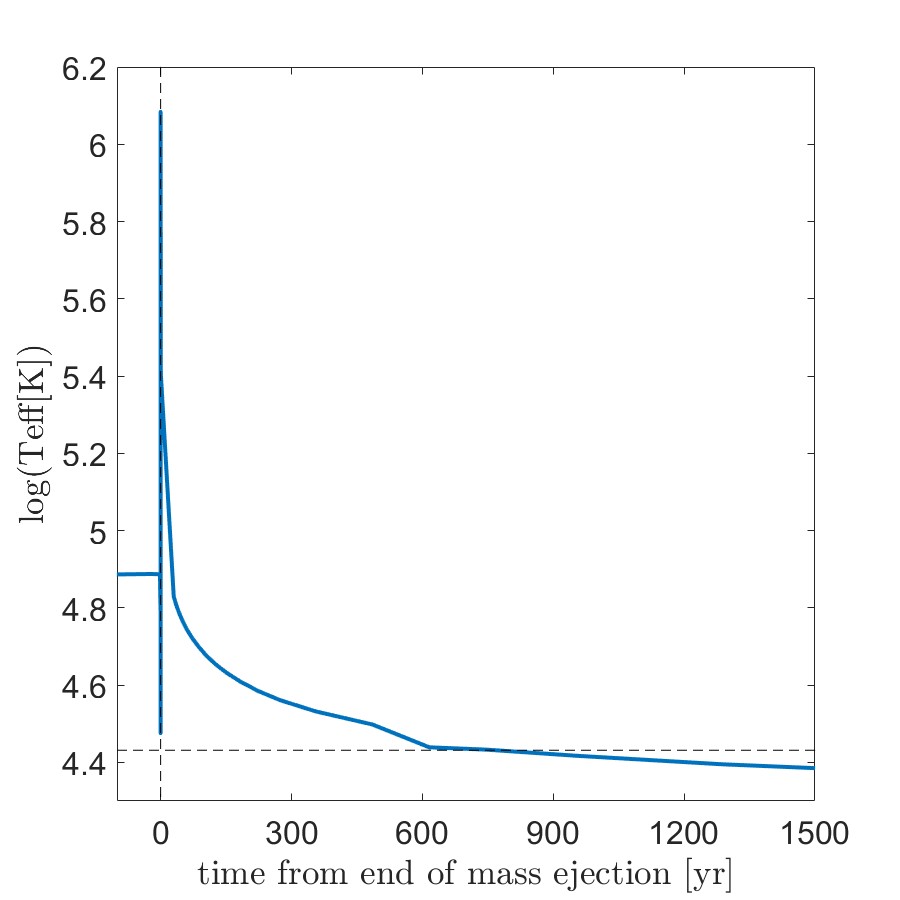}
    \caption{The effective temperature of the photosphere of the white dwarf in U Gem after one of the nova eruptions of Figure 2, as predicted by the models of Section 2. The coolest temperature observed for the white dwarf is $\sim$ 27,000 Kelvin, indicated by the dashed horizontal line, which corresponds to a cooling age $\sim$ 800 yrs since the last nova eruption. }
\end{figure*}

\begin{figure*}[h!]
    \centering
    \includegraphics[width=7in]{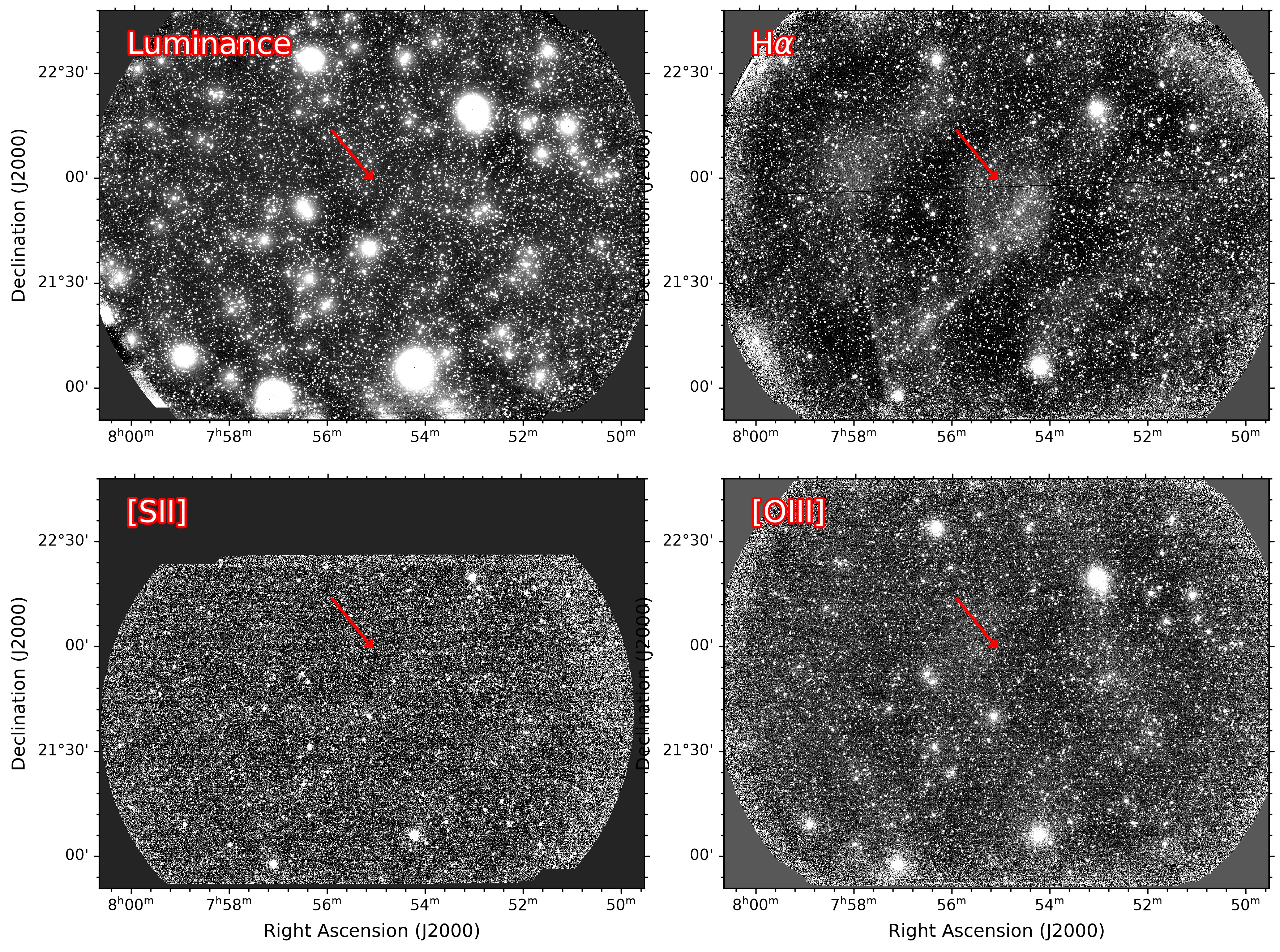}
    \caption{Condor Array Telescope images in 4 passbands (indicated in the top left of each figure) of the field of U Gem. Exposure times are given in Table 1. The dwarf nova is at the tip of a red arrow, which indicates U Gem's Gaia-derived proper motion over the past 25,000 years. A nearly horizontal line near the middle of the H$\alpha$ image is an artifact.}
\end{figure*}

\newpage

\begin{figure*}[h!]
    \centering
    \includegraphics[width=7in]{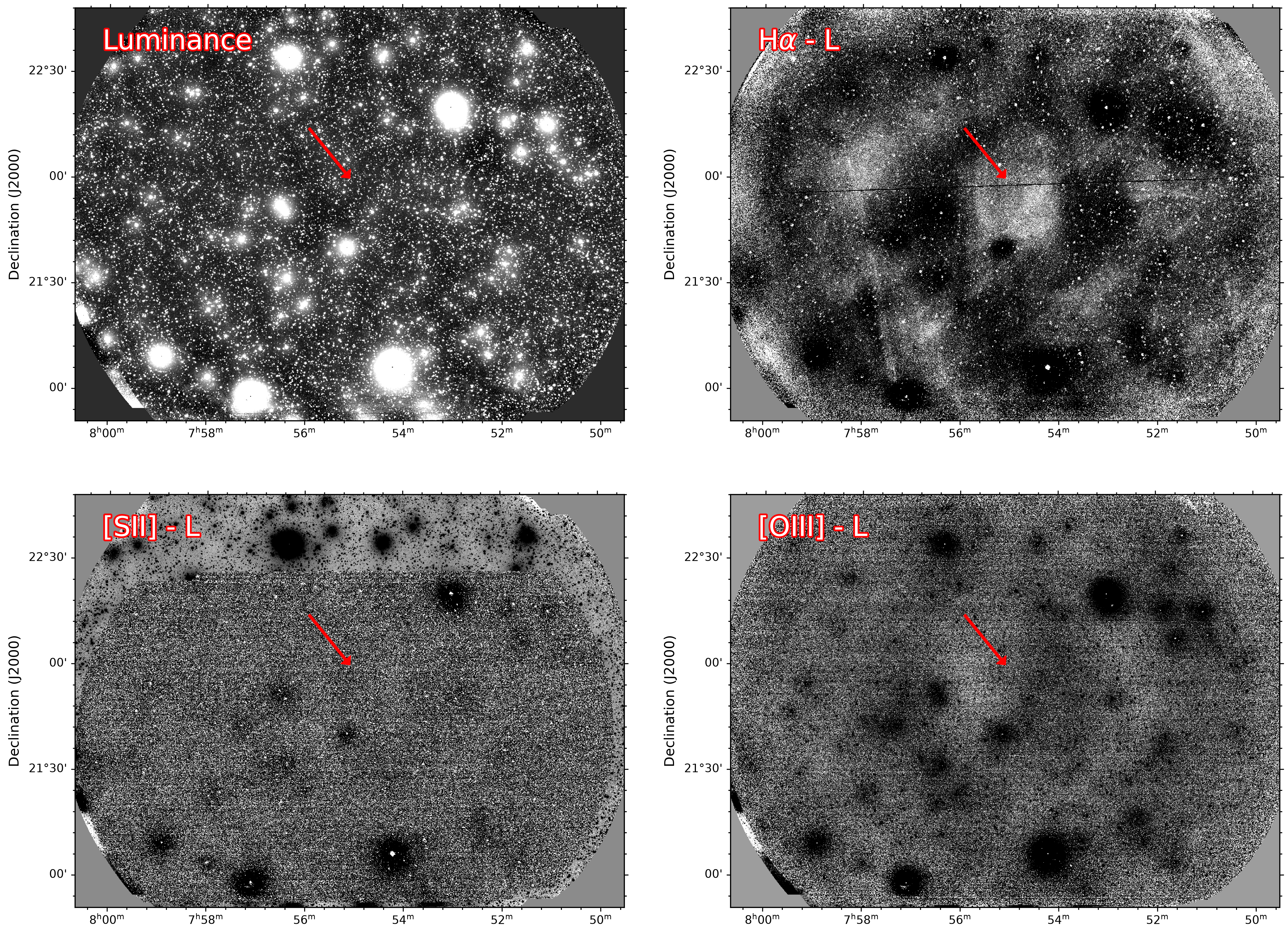}
    \caption{Difference images of U Gem (position indicated with a red arrow) created by directly subtracting the Luminance coadded image from each of the three narrowband coadded images of Figure 4, then applying 8x8 block-summing. All images are scaled uniformly to allow direct comparison of the faint emission features. A nearly horizontal line near the middle of the H$\alpha$ image is an artifact.}

\end{figure*}

\newpage
\begin{figure*}[h!]
    \centering
    \includegraphics[width=7in]{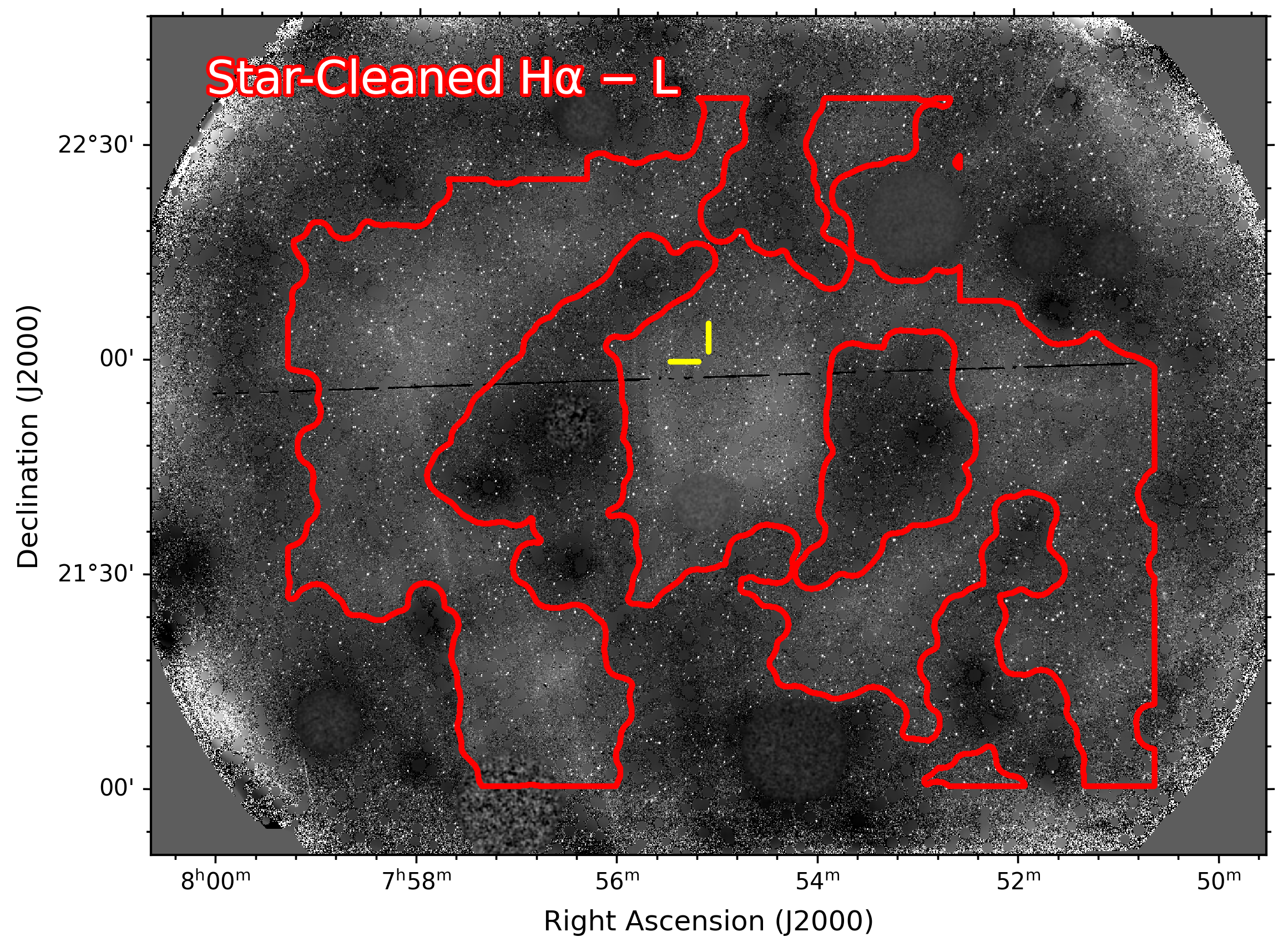}
    \caption{Difference image (H$\alpha$-L) of U Gem after removal of stars, with a constant-brightness contour superposed. A central nebulosity is flanked to the East and West by nearly equal-brightness arcs of nebulosity. The location of U Gem is indicated by yellow tick marks.}
\end{figure*}

\newpage
\begin{figure*}[h!]
    \centering
    \includegraphics[width=7in]{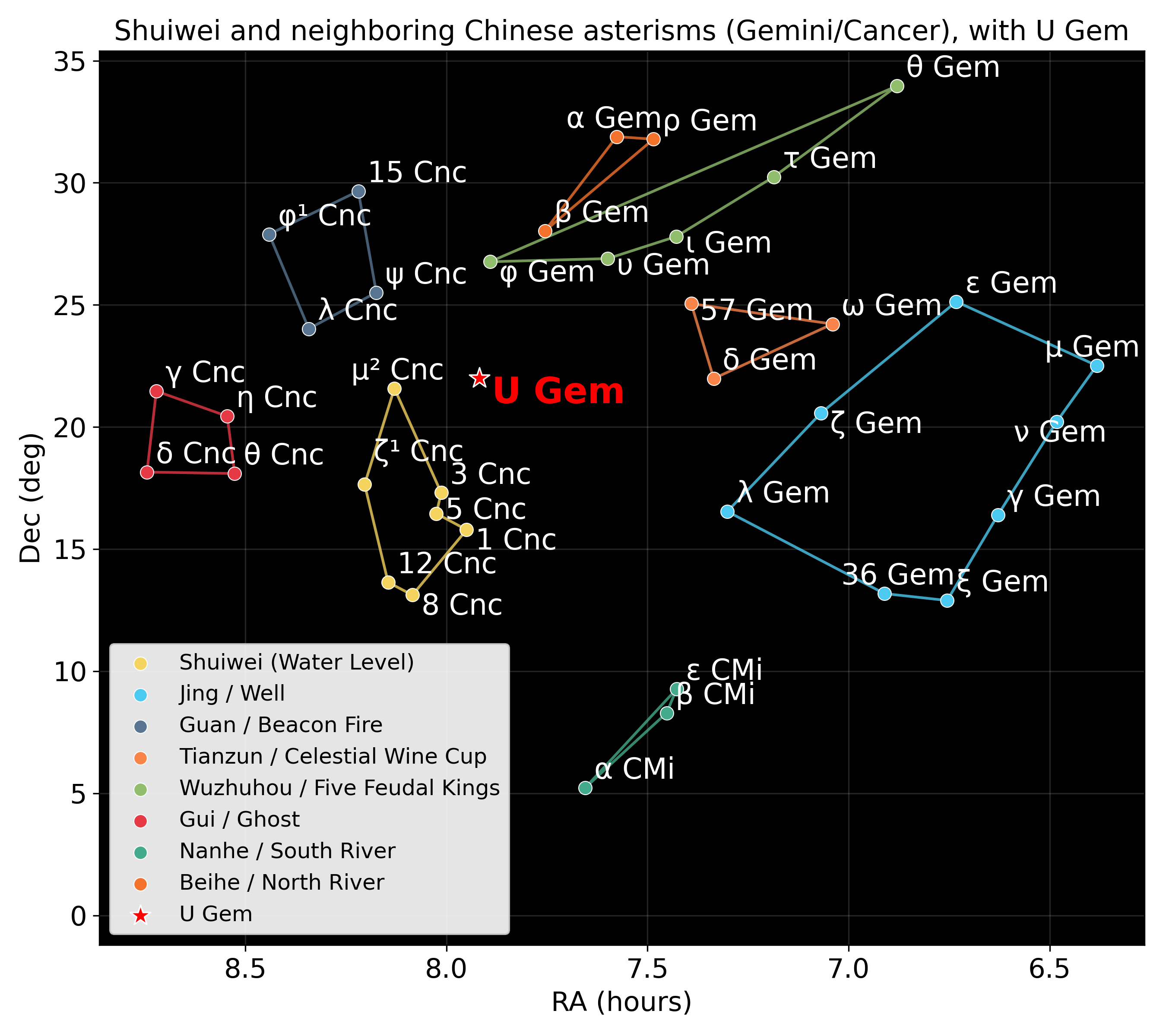}
    \caption{The position of U Gem relative to its surrounding Chinese asterisms. U Gem is closest to the Shuiwei (Water Level) asterism.}
\end{figure*}

\end{document}